\documentstyle[multicol,prl,aps,graphics]{revtex}
\begin{document}
\title{Metal-insulator transitions:\\Influence of lattice structure,
Jahn-Teller effect, and Hund's rule coupling}  

\author{J.E. Han, E. Koch and O. Gunnarsson}   
\address{Max-Planck-Institut f\"ur Festk\"orperforschung, 
D-70506 Stuttgart, Germany}

\date{\today}
\maketitle
\begin{abstract}
We study the influence of the lattice structure,  the Jahn-Teller
effect and the Hund's rule coupling on a   metal-insulator
transition in A$_n$C$_{60}$ (A= K, Rb). The difference in 
lattice structure favors A$_3$C$_{60}$ (fcc) being a metal and 
A$_4$C$_{60}$ (bct) being an insulator, 
and the coupling to H$_g$ Jahn-Teller phonons favors A$_4$C$_{60}$ being 
nonmagnetic. The coupling to H$_g$ (A$_g$)  
phonons decreases (increases) the value $U_c$ of the Coulomb integral  
at which the metal-insulator transition occurs. There is an important partial 
cancellation between the Jahn-Teller effect and the Hund's rule coupling.
\end{abstract}
\begin{multicols}{2}
The competition between the Coulomb repulsion, the kinetic energy, 
the Jahn-Teller effect and the Hund's rule coupling leads to 
interesting physics. Examples  
are perovskites, e.g., the manganites\cite{Imada}, 
and alkali-doped fullerenes\cite{RMP}.                                 
Here we focus on the metal-insulator transition for an integer 
number of electrons per site. This is particularly relevant
for the fullerenes, since A$_3$C$_{60}$ (A= K, Rb) is a 
metal\cite{metal} while A$_4$C$_{60}$ is a nonmagnetic 
insulator\cite{insulator,nonmagnetic}. According to band theory 
both are  metals\cite{Erwin}, and A$_4$C$_{60}$ must 
therefore be an insulator due to interactions left out in band 
structure calculations. 

The metal insulator transition in a correlated system is usually
discussed in terms of the ratio $U/W$\cite{Georges}, where
$U$ is the Coulomb interaction between two electrons on the same 
molecule and $W$ is the one-particle band width $W$. The ratio 
$U/W$ is, however, almost identical for A$_3$C$_{60}$ and 
A$_4$C$_{60}$\cite{Erwin,RMP,A4}. The question is then why not both            
systems are either metals or insulators. To study this, we apply 
the dynamical mean-field theory (DMFT), projection Quantum 
Monte-Carlo (QMC) and exact diagonalization techniques to models 
of A$_n$C$_{60}$. 

For the Fullerenes it is believed that  $U/W\sim 1.5-2.5$\cite{RMP}.
In spite of this large ratio, these systems are close to a 
metal-insulator transition due to the orbital degeneracy $N=3$ 
of the partly occupied $t_{1u}$ band\cite{deg,Jong}.                           
The lattice structure is fcc for A$_3$C$_{60}$ and bct for A$_4$C$_{60}$.
The important electron-phonon coupling is to H$_g$ Jahn-Teller phonons.
We find that the difference in lattice structure alone can explain why 
A$_3$C$_{60}$ is a metal but A$_4$C$_{60}$ is an insulator and that the 
electron-phonon coupling can explain why  A$_4$C$_{60}$ is  
nonmagnetic. We find an important competition between the 
Jahn-Teller effect and the Hund's rule coupling.  The H$_g$ and A$_g$ 
intramolecular phonons are found to have the opposite effect on the 
critical $U_c$, for which the metal-insulator transition occurs.

We consider a model of A$_n$C$_{60}$ which includes a three-fold degenerate
$t_{1u}$  level on each molecule and the hopping between 
different molecules 
\begin{equation}\label{eq:1}
H_{\rm hop}=\sum_{\sigma,m}\varepsilon_{t_{1u}}
n_{i\sigma m}+\sum_{<ij>\sigma mm'}
t_{ijmm'}\psi^{\dagger}_{i\sigma m} \psi_ {j\sigma m'}, 
\end{equation}
where $\psi^{\dagger}_{i\sigma m}$ creates an electron on 
molecule $i$ with the quantum number $m$ and spin $\sigma$.
The hopping matrix elements $t_{ijmm'}$\cite{TB}  include the 
orientational disorder\cite{Stephens} and the lattice     
structure, with nearest neighbor hopping for the fcc structure
and a weak second nearest neighbor hopping for the bct 
structure\cite{A4}. The Coulomb interaction is given by   
\begin{eqnarray}\label{eq:2}
H_{\rm U}=&&U_{xx}\sum_{im} n_{im\uparrow}n_{im\downarrow}
+U_{xy}\sum_{i\sigma\sigma^{'}}\sum_{m<   m^{'}}n_{i\sigma m}
n_{i\sigma^{'}m^{'}}  \nonumber  \\
+&&{1\over 2}K \sum_{i\sigma\sigma^{'}}\sum_{m\ne m^{'}}
\psi^{\dagger}_{im\sigma}\psi^{\dagger}_{im^{'}\sigma^{'}}
\psi_{im\sigma^{'}}\psi_{im^{'}\sigma }   \\
+&&{1\over 2}K \sum_{\sigma}\sum_{m\ne m^{'}}
\psi^{\dagger}_{m\sigma }\psi^{\dagger}_{m-\sigma }
\psi_{m^{'}-\sigma }\psi_{m^{'}\sigma },  \nonumber  
\end{eqnarray}
where $U_{xx}$ and $U_{xy}$ describe the interaction between equal 
and unequal orbitals, respectively. $K$ is 
an exchange integral and $U_{xx}=U_{xy}+2K$. Finally we include the
interaction with a five-fold degenerate H$_g$ phonon on each site
$$H_{\rm ph}= 
\omega_{ph}\sum_{i\nu}b^{\dagger}_{i\nu}b_{i\nu} 
+ {g\over 2} \sum
_{i\nu\sigma m m^{'}}
V_{mm^{'}}^{(\nu)}c^{\dagger}_{im\sigma} c_{im^{'}\sigma}(b_{i\nu}+ 
b_{i\nu}^{\dagger}),$$
where $b_{i\nu}$ creates a phonon with the quantum number $\nu$
on the molecule $i$. The matrices 
$V_{mm^{'}}^{(\nu)}$ are determined by symmetry\cite{c60jt}. 
The coupling constant $g$ is related to the dimensionless 
electron-phonon coupling $\lambda=(5/3)N(0)g^2/\omega_{ph}$.                                 
We also consider the coupling to A$_g$ phonons,
for which $V^{\nu}_{mm^{'}}$ is diagonal in $m$ and $m^{'}$.

\begin{figure}[bt]
\centerline{
\rotatebox{270}{\resizebox{!}{2.3in}{\includegraphics{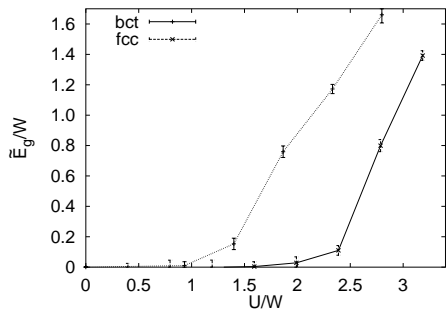}}}}
\caption[]{\label{fig1} The energy gap
$\tilde E_g$ (Eq. (\ref{eq:5})) as a function of $U/W$ for 
$M=32$ molecules on fcc and bct lattices and the filling $n=4$.
The lines are guides for the eye. }
\end{figure}

In a first step we analyze the effect of the lattice structure
alone, neglecting the electron-phonon coupling ($g=0$) and the 
multiplet effects ($K=0$ and $U_{xx}=U_{xy}\equiv U$). 
We use a projection Quantum Monte-Carlo (QMC) $T=0$ method
in the fixed node approximation\cite{tenhaaf}, which gives 
quite accurate ground-state results for this model\cite{deg}.
A$_3$C$_{60}$ and A$_4$C$_{60}$ differ in the number $n$ of 
conduction electrons per site and in the lattice structures. 
For a fcc lattice, $n=3$ and $n=4$ give Mott transitions at almost 
the same $U_c$\cite{filling}. We therefore focus on the difference in 
lattice structure, and consider $n=4$ for clusters with $M$ molecules 
put on fcc or bct lattices. The band gap for filling $n$ is
\begin{equation}\label{eq:4}
E_g=E(nM+1)+E(nM-1)-2E(nM),
\end{equation}
where $E(N)$ is the energy of a system with $N$ electrons. We want
to extrapolate to $M\to \infty$ and determine the $U_c$ for
which $E_g$ is zero. To reduce the finite size effects\cite{deg},
we add
\begin{equation}\label{eq:5}
\tilde E_g(U)=E_g(U)-{U\over M}-E_g(U=0),
\end{equation}
where $E_g(U=0)$ is the band gap for $U=0$. These corrections go 
to zero for large $M$, but they improve the extrapolation 
$M\to \infty$. Fig. \ref{fig1} shows that          
the metal-insulator transition happens for a substantially smaller
$U/W$ for the bct ($U_c/W\sim 1.3$) than for the 
fcc structure ($U_c/W\sim 2.3 $). The insulating state is 
antiferromagnetic.

To understand these results, we note that on the fcc lattice
it is possible to hop on a triangle, i.e., to return to the
original site after three hops. On a bct lattice, on the other hand, 
this is not possible if the small second nearest neighbor hopping 
integrals are neglected. The simplest systems with these properties 
are a triangle and a square, each site having a level with spin but 
no orbital degeneracy. A nearest neighbor hopping integral $t<0$ connects 
the orbitals. The one-particle spectrum is $\pm2 t$ for the square and
$-2|t|$ and $t$ for the triangle. For the triangle there is a state 
with maximum bonding character ($-2|t|$), but it is not possible     
to construct an optimally anti-bonding state, due to the presence of 
frustration. Thus the one-particle band width  are $W=3t$ 
and $4t$ for the triangle and the square, respectively. The curves in Fig.
\ref{fig1} mainly differ in the large $U$ limit and we therefore
consider this limit. We construct the many-body states
of the triangle with two, three and four electrons, which determine the 
band gap (Eq.~(\ref{eq:4})). The energy $E(3)=O(t^2/U)$,
since hopping is suppressed to order $t/U$.
For the case of four electrons, we construct all states with the minimum
(one) double occupancy and $S_z=0$. These states describe how the 
double occupancy hops around the triangle. The original state is, 
however, not recovered after one loop, since the spins on the sites with
a single occupancy have been flipped. Moving the double occupancy    
around the triangle a second time 
restores the spins and the original state is recovered 
after six moves. The corresponding $6\times 6$ matrix has the extreme 
eigenvalues $\pm 2t$. In the lowest many-body state of the triangle
with four electrons, it is therefore not possible to restore the state in an 
odd number of hops, and the frustration does not show up. In a similar 
way we obtain the lowest energy $-2t$ for the two-electron
state. The square has the same energies. Thus 
\begin{eqnarray}\label{eq:6}
\begin{array}{ll}
E_g=U-4|t|=U-{4\over 3}W & \mbox{for a triangle} \\ 
E_g=U-4|t|=U-W & \mbox{for a square}  
\end{array}
\end{eqnarray}
Both the triangle and the square have no frustration in their many-body 
states, and for fixed $t$ the gaps are the same.
The one-particle band width $W$, however, is reduced by the frustration
in the triangle, and expressing the $E_g$ in terms of $W$ 
requires a larger prefactor in the frustrated case. 
These results give a qualitative explanation of Fig. \ref{fig1}.

Although the calculation above can explain why A$_4$C$_{60}$ 
is an insulator while A$_3$C$_{60}$ a metal, it incorrectly 
predicts A$_4$C$_{60}$ to be antiferromagnetic. The  calculation 
neglects, however, the coupling to the Jahn-Teller phonons,
which tends to make A$_4$C$_{60}$ a nonmagnetic insulator\cite{Fabrizio}.
The electron-phonon interaction has been estimated from photoemission 
experiments for a free molecule\cite{PES}. We describe the eight H$_g$
phonons by an effective mode, with the logarithmically averaged 
frequency $\omega_{ph}=0.089$ eV, and the effective coupling  
$g=0.089$ eV. For a free molecule this leads to a singlet being 
0.29 eV below the lowest triplet. This triplet-singlet splitting is 
larger than an experimental estimate of 0.1 eV for 
A$_4$C$_{60}$\cite{nonmagnetic}. The splitting is, however, reduced 
by the competition with the Hund's rule coupling.  An estimate of 
the exchange integral $K$ based
on an {\it ab initio} SCF calculation\cite{correlated} gave $K=0.11$  
eV\cite{K}.  This number is, however, expected to be reduced by 
correlation effects. For instance, for atomic multiplets a reduction
by 25 $\%$ has been found\cite{Barth}. Indeed, we find that the experimental
triplet-singlet splitting is reproduced by using $K=0.07$ eV.

Since the metal-insulator transition depends on a competition between the
kinetic and Coulomb energies, and since we may expect the 
electron-phonon coupling to reduce the hopping, we may expect this
to reduce $U_c$\cite{Tosatti}.  We therefore study the effect of 
phonons on $U_c$ (for $K=0$).

For this purpose we apply the dynamical-mean field theory 
DMFT\cite{Georges}. We use hopping integrals for a Bethe lattice 
in the infinite 
dimensional limit $t_{imjm^{'}}\sim t^{\ast}\delta_{mm^{'}}/\sqrt{z}$, 
where $z\to \infty$ is the connectivity. The impurity model, resulting 
in the DMFT, is solved with a QMC method\cite{Hirsch}. The phonon 
fields are treated fully quantum mechanically, and they are updated 
together with the Fermion auxiliary fields in each Monte Carlo 
step\cite{Scalapino}. We use the one-particle band width $W=2$ and a 
Trotter break up $\Delta \tau=1/3$.  

For an insulator $G(\tau=\beta/2)$ decays
exponentially with $\beta$, where $G(\tau)$ is the electron Green's
function on the imaginary time axis. We therefore use $G(\beta/2)$
to determine whether the system is a metal or an insulator.

\begin{figure}[bt]
\centerline{
\rotatebox{270}{\resizebox{!}{2.5in}{\includegraphics{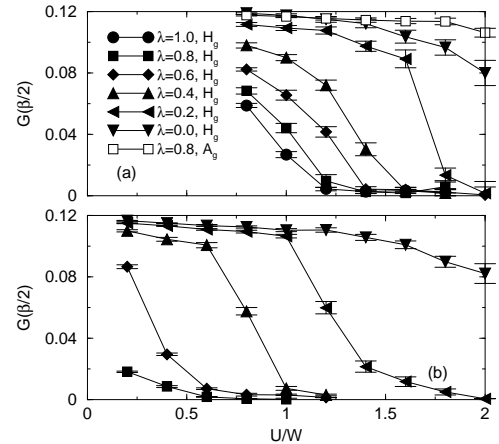}}}}
\caption[]{\label{fig2}(a) The electron Green's function $G(\tau=\beta/2)$ 
as a function of $U/W$ for the filling $n=3$ and different values of the 
electron-phonon coupling $\lambda$. The figures compares the coupling 
to A$_g$ and H$_g$ phonons. (b) $G(\beta/2)$ for coupling to H$_g$
phonons and $n=4$.  $G(\beta/2)\approx 0$ implies an insulator. 
}
\end{figure}

We first compare the coupling to A$_g$ and H$_g$ phonons for  $n=3$. 
Fig. \ref{fig2}a shows that $G(\beta/2)$ is reduced as $U/W$ is
increased, since the system gets closer to a
metal-insulator transition. For $\lambda=0$ extrapolation suggests
a rather large $U_c/W$.  For H$_g$ phonons an increase in $\lambda$ leads to 
a rapid reduction of $G(\beta/2)$ and $U_c$, while 
for $A_g$ phonons this leads to an {\it increase} in $G(\beta/2)$ and $U_c$.

To understand these results we study a free molecule 
(Table \ref{table1}) and a system consisting of two molecules 
(dimer) (Table \ref{table2}) in the limit
\begin{equation}\label{eq:7}
K\sim {g^2\over \omega_{ph}}\equiv E_{JT}\ll \omega_{ph}\ll W\ll U.
\end{equation}
 
Table \ref{table2} shows the energy gap of the dimer.
In agreement with the full DMFT results ($K=0$
and $n=3$) the gap is increased by a coupling to H$_g$ 
but decreased by a coupling to A$_g$ phonons. We first 
consider the A$_g$ case. Since $V_{mm^{'}}=\delta_{mm^{'}}V_{mm}$ we can 
transform the electron-phonon coupling to the form 
\begin{equation}\label{eq:8}
g\sum_i (n_i-n)(b_i+b_i^{\dagger}),
\end{equation}
where $n_i$ is the total occupation number operator for site $i$
and $n$ is the (integer) filling. An irrelevant constant has been neglected. 
We first study    the state  with $2n$ electrons. In the limit $W\ll U$ 
hopping is suppressed, and $n_i-n\approx 0$.
The coupling (Eq. (\ref{eq:8})) is then negligible, and the 
electron-phonon contribution to the energy is small. In the case 
of an extra electron or hole, however, this additional charge can 
hop even for $W\ll U$.  The coupling to the phonons then 
lowers the energy, and according to Eq. (\ref{eq:4}) this reduces the gap.
 
For coupling to H$_g$ phonons, the state with $2n$ electrons can 
lower its energy via the (dynamic) Jahn-Teller effect. Since hopping is 
very efficiently suppressed, the energy gain is accurately given as 
twice the energy for a free molecule (Table \ref{table1}). In the case of 
an extra electron or hole, on the other hand, hopping dominates over     
the Jahn-Teller effect in the limit (\ref{eq:7}). The system can 
then only take advantage of this effect to the extent 
that it does not interfere with the hopping. The electron-phonon
coupling then gives a much  smaller lowering of the energy than for 
the state with $2n$ electrons, which increases the gap (Eq. (\ref{eq:4})).

Fig. \ref{fig2}b shows results for coupling to H$_g$ phonons
and filling $n=4$. $U_c/W$ is smaller than for $n=3$, although the 
lattice structure is the same as for $n=3$. This can be 
understood from Table \ref{table1}, which shows that the energy gain 
in the free molecule due to the electron-phonon coupling is
larger for $n=4$. This enters in $E(nM)$, while the electron-phonon
coupling plays a smaller role for $E(nM\pm 1)$. The electron-phonon 
coupling {\it alone} would then tend to favor A$_4$C$_{60}$ being an insulator 
and A$_3$C$_{60}$ being a metal. As we will see below, this effect is,    
however, partly cancelled by the Hund's rule coupling.

The coupling to the H$_g$ phonons pushes $U_c$ for A$_3$C$_{60}$
to the lower end of the physical range of $U/W$, raising some
questions of why not also A$_3$C$_{60}$ is an insulator. Although, 
the A$_g$ phonons tends to increase $U_c$, this should not be important
due to the weak coupling to the A$_g$ phonons\cite{screen}.
However, there is a substantial coupling to a plasmon in 
A$_3$C$_{60}$\cite{plasmon}. This should tend to increase $U_c$, 
since it couples to the electrons in the same way as the A$_g$ 
phonons. Below we show that the Hund's rule coupling also
plays an important role in this context.

\begin{figure}[bt]
\centerline{
\rotatebox{270}{\resizebox{!}{2.2in}{\includegraphics{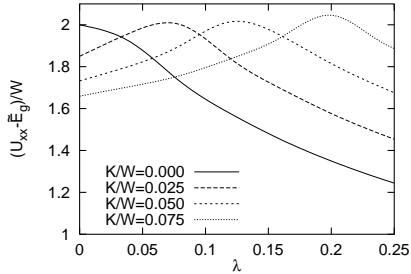}}}}
\caption[]{\label{fig4}The estimate $U_{xx}-\tilde E_g$ of the critical 
$U_{xx}$ as a function of $\lambda=4g^2/(\omega_{ph}W)$ for 
different values of $K$ and $U_{xx}/W=3$. }
\end{figure}

We next consider the effects of the Hund's rule coupling ($K>0$).
Since these terms in Eq. (\ref{eq:2}) lead to a sign-problem in 
the DMFT QMC calculation,
we use exact diagonalization. To reduce the size of the Hilbert
space we consider a four-site system with  two-fold orbital and phonon
degeneracies. The nearest neighbor hopping 
$t_{im,jm^{'}}=t_{ij}\delta_ {mm^{'}}$ is chosen randomly, thus reducing
the degeneracy and the one-particle spacing. We limit the size of the
Hilbert space by allowing a maximum of two phonons per site. 
Due to this limitation, the calculation is not fully converged
for the larger coupling constants considered below. From the finite size
corrected band gap $\tilde E_g(U_{xx})$ we estimate the critical 
$U_{xx}$ as $U_{xx}-\tilde E_g(U_{xx})$, shown in Fig. \ref{fig4}.
The figure illustrates that for $\lambda=0$ an increase in $K$ leads
to a decrease in $U_c$\cite{Jong}. In analogy to the discussion
for the Jahn-Teller effect, the Hund's rule coupling can effectively
lower the energy of the state with $nM$ electrons while for the 
states with $nM\pm 1$ electrons, the stronger interference with
hopping leads to a smaller lowering of the energy. For $\lambda>0$    
the competition between the Jahn-Teller effect and the Hund's rule coupling  
tends to reduce the influence of either effect on $U_c$. This is shown in 
Table \ref{table1} and \ref{table2} and in Fig. \ref{fig4}.

To summarize, we have found that the difference in lattice structure
favors A$_3$C$_{60}$ being a metal and A$_4$C$_{60}$ being an
insulator. The Jahn-Teller effect wins over the 
Hund's rule coupling, making A$_4$C$_{60}$ a nonmagnetic 
insulator.  The coupling to the H$_g$ phonons tends to strongly reduce  
the critical $U$ for a metal-insulator transition, raising questions
about why not also A$_3$C$_{60}$ is an insulator. This effect is, however,
partially cancelled by the Hund's rule coupling. 
The coupling to plasmons tends to further increase the 
critical $U$.

This work has been supported by the Max-Planck-Forschungspreis.

\begin{table}
\caption[]{The ground-state energy $E(n)$ of an isolated molecule
for $n$ electrons. The quantity $n(n-1)U_{xy}/2$ has been subtracted.
The results are symmetric around $n=3$.}
\begin{tabular}{ccc}
\multicolumn{1}{c}{$n$} &
\multicolumn{2}{c}{$\tilde E(N)\equiv E(N)-N(N-1)U_{xy}/2$}  \\
    &  \rm Low spin ($K<{3\over 2}E_{JT}$)   &  High spin 
($K\ge {3\over 2}E_{JT}$)  \\
\tableline
   1    &\multicolumn{2}{c}{ $-{5\over 2}E_{JT}$}                 \\
   2    & $-10E_{JT}+4K     $  &  $ -{5\over 2}E_{JT}-K$  \\
   3    & $-{15\over 2}E_{JT}+2K$ &$ -3K$                 \\
\end{tabular}\label{table1}
\end{table}

\begin{table}
\caption[]{$E_g(n)-U_{xy}-d_3(n)t$ for a two-site 
model as a function of the filling $n$. The hopping contribution to the 
gap is $d(n)t$, where $d_3(n)=$ -3, -5 and -6 for $n=$ 1, 2 and 3,
respectively. The coupling is to one $A_g$ (A) or one H$_g$
(H) phonon per site. The results are symmetric around $n=3$.}
\begin{tabular}{ccccc}
\multicolumn{1}{c}{Phonon}  &
\multicolumn{1}{c}{$n$} & \multicolumn{3}{c}{$E_g(n)-U_{xy}-d_3(n)t$} \\
\tableline
& & $K\le {3\over 2}E_{JT}$ & ${3\over 2}E_{JT}< K \le {9\over 4}E_{JT}$
& $K>{9\over 4}E_{JT}$ \\
\tableline
A&1 & \multicolumn{3}{c}{$-E_{JT}- K$} \\
A&2 & \multicolumn{3}{c}{$-E_{JT}+3K$}  \\  
A&3 & \multicolumn{3}{c}{$-E_{JT}+12K$}       \\                          
H&1 & \multicolumn{2}{c}{$5E_{JT}+{2\over 3}K$} & ${35\over 4}E_{JT}- K$ \\
H&2 & $35E_{JT}-{46\over 3}K$ & $ 5E_{JT}+
{14\over 3}K $ & $ {35\over 4}E_{JT}+3K$ \\
H&3 & ${55\over 2}E_{JT}-8K$ & \multicolumn{2}{c}{
$-{5\over 2}E_{JT}+12 K$}\\ 
\end{tabular}\label{table2}
\end{table} 
\end{multicols}
\end{document}